\documentclass[amsmath,amssymb,reprint,aps,pra,floatfix,superscriptaddress]{revtex4-1}
\usepackage{bm}
\usepackage{amsmath}
\usepackage{graphicx}
\graphicspath{ {./Figures/}, {./SM/} }
\usepackage{bm}
\DeclareMathOperator{\Tr}{Tr}
\usepackage{braket}
\usepackage{epstopdf}
\usepackage{bm}
\usepackage{bbm}
\usepackage{lineno}
\usepackage{hyperref}
\usepackage[title]{appendix}
\usepackage{etoolbox}
\patchcmd{\appendices}{\quad}{: }{}{}

\begin{document}

\title{Drive-noise tolerant optical switching inspired by composite pulses}

\author{J.F.F. Bulmer}\email{jacob.bulmer@bristol.ac.uk}
\affiliation{Quantum Engineering Technology Labs, University of Bristol, Bristol BS8 1FD, UK}
\affiliation{Clarendon Laboratory, University of Oxford, Parks Road, Oxford, OX1 3PU, UK}

\author{J.A. Jones}
\affiliation{Clarendon Laboratory, University of Oxford, Parks Road, Oxford, OX1 3PU, UK}

\author{I.A. Walmsley}
\affiliation{Clarendon Laboratory, University of Oxford, Parks Road, Oxford, OX1 3PU, UK}
\affiliation{Department of Physics, Imperial College London, London SW7 2AZ, UK}

\begin{abstract}
Electro-optic modulators within Mach--Zehnder interferometers are a common construction for optical switches in integrated photonics.
A challenge faced when operating at high switching speeds is that noise from the electronic drive signals will effect switching performance.
Inspired by the Mach--Zehnder lattice switching devices of Van Campenhout \textit{et al.} [Opt. Express, 17, 23793 (2009)]\nocite{Campenhout:09} and techniques from the field of Nuclear Magnetic Resonance known as composite pulses, we present switches which offer protection against drive-noise in both the on and off state of the switch for both the phase and intensity information encoded in the switched optical mode.
\end{abstract}

\maketitle

\section{Introduction}
Optical switching is a key tool in many areas of integrated photonics.
For example, it is thought that optical switching may be useful in removing  inter-chip~\cite{4509424,sun-single-chip-2015,5071309} and intra-chip~\cite{Cheng:18} communication bottlenecks in future classical computing architectures.
Optical switching also finds applications in many proposed platforms for quantum computing. Several architectures use optical switching networks for connecting matter based quantum bits (qubits) via single photon entangling gates, allowing for distributed architectures~\cite{PhysRevA.89.022317,PhysRevX.4.041041,choi2019percolation}.
Linear optical quantum computing (LOQC) uses single photons for both transportation and processing of quantum information~\cite{knill2001scheme,RevModPhys.79.135,Rudolph17}.
In LOQC, optical switching networks are required for multiplexing of nondeterministic, heralded quantum processes to near determinism, and act on both single photons and multi-photon entangled states~\cite{Bonneau15,PhysRevX.5.041007,mercedes-thesis:15}.
Measurement based LOQC architectures also require fast optical switching to adaptively set the basis of single qubit measurements.
A popular platform for these technologies is silicon photonics, which offers high component density and compatibility with CMOS fabrication~\cite{7479523}. However, a wide variety of other materials are also being developed, such as lithium niobate \cite{wang2018integrated}, silicon nitride \cite{alexander2018nanophotonic}, silica \cite{Carolan711}, III-V semiconductor devices~\cite{Stabile:12}, and hybrid combinations thereof~\cite{he2019high, Li:18}.
For this reason we will take an abstracted, hardware platform agnostic approach in this work.

\begin{figure}
\centering\includegraphics[width=0.45\textwidth]{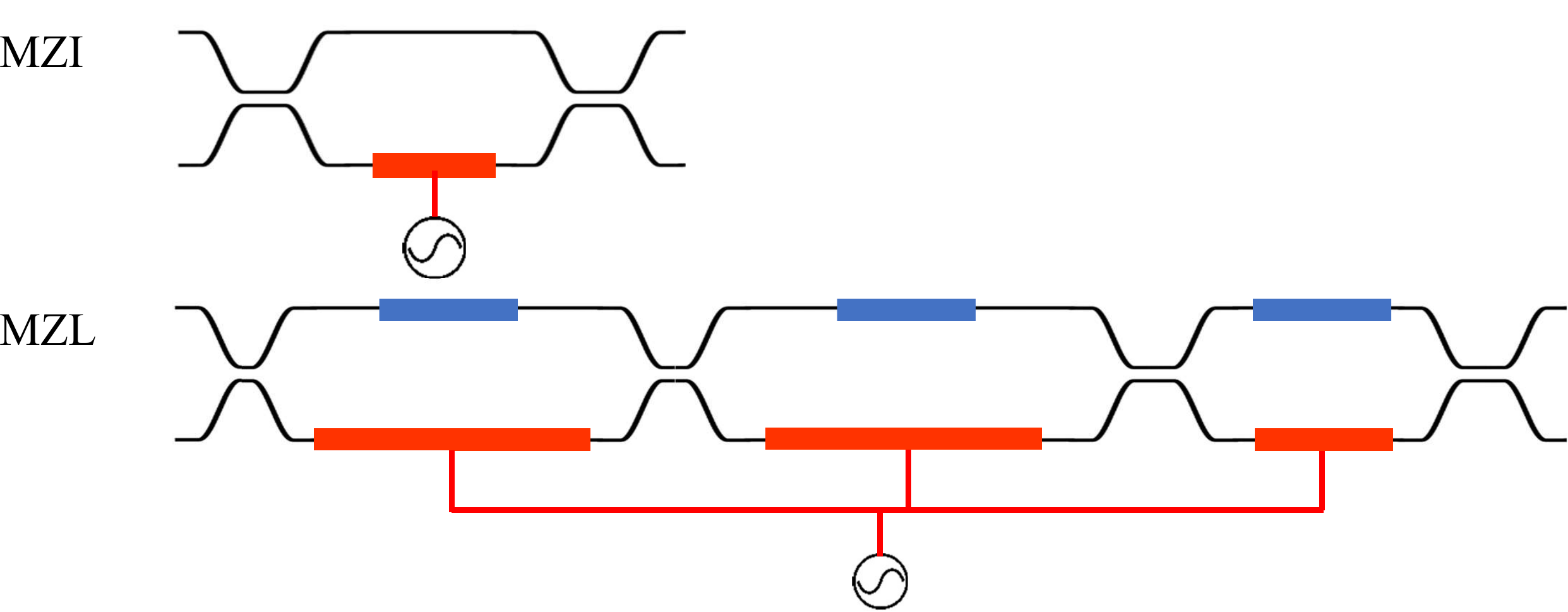}
    \caption{Schematic for the layout of an MZI and an MZL. The MZI has a single modulator, shown as a red rectangle. In the MZL we have multiple modulators, all controlled by the same drive signal. The modulators can have different lengths and so can impart different but proportional phase shifts. Blue rectangles imply independent and static programmable phase shifts.}
    \label{fig1}
\end{figure}

An electro-optic modulator within a Mach--Zehnder interferometer (MZI) is a typical construction for an optical switch.
Because MZIs do not rely on moving parts or resonance effects, such as for microelectromechanical systems (MEMS) or ring resonator based switches, they offer reliable, high speed operation and good thermal stability.
However, any noise in the signal which drives the modulator will cause imperfect switching, which leads to optical crosstalk.
Reducing this crosstalk is important for many applications.
To keep communication errors in optical switching networks constant, optical crosstalk must be lower than the worst case loss~\cite{4688609}.
As the complexity of optical switching networks increases, the worst case loss of the switch will increase, meaning that the allowable crosstalk for each switch decreases. 
Whilst intelligently designed driving electronics can reduce the noise of the drive signals, commonly occurring issues such as interference from switch mode power supplies (for example those used for temperature stabilisation) and intersymbol interference remain challenging to suppress.
Because of these kinds of engineering challenges, noise in the electrical signals which drive electro-optic modulators has been identified as a key issue in realising large scale optical switching networks for classical computing applications~\cite{Campenhout:09}.
In quantum communication schemes, intersymbol interference effects have been highlighted as a potential security flaw in integrated photonic quantum communication systems~\cite{10.1117/12.2307409}.
Drive-noise is also identified as the expected dominant source of stochastic noise in LOQC~\cite{Rudolph17}. 
In quantum error correction codes, the error thresholds depend on both stochastic error and photon loss.
Importantly, this means that any improvement to stochastic error rates will allow for more relaxed requirements on photon loss~\cite{PhysRevLett.105.200502}.

In an ideal perfectly balanced MZI, the phase shift imparted on the switched optical modes does not depend on the phase shift imparted by the modulator.
This is not true in general when unbalanced couplers are used.
Many applications of optical switching use the phase of the switched optical modes to carry information, such as phase-shift keying communication schemes~\cite{1303576} and multiplexing of entangled photonic state generation. Therefore, such phase shifts from an optical switch can be problematic.
However, for applications where the phase of an optical mode does not carry information, for example in pulse amplitude modulation communications schemes and photon source multiplexing, a noisy phase shift has no impact on performance.
This application dependent difference motivates us to use two fidelity measures (defined in section~\ref{model}) to assess the performance of our devices, one which is sensitive to phase errors and one which is not.

In an effort to solve the issues caused by drive-noise, designs have been proposed~\cite{Campenhout:09} and realised~\cite{VanCampenhout:11} for multi-stage Mach--Zehnder interferometers, known as a Mach--Zehnder lattices (MZL).
These devices exhibit excellent drive-noise tolerance in their on state when assessed using phase insensitive measures. 
An important part of the design of these devices is that the modulation at every stage is controlled by the same drive signal, as depicted in Figure~\ref{fig1}. 
One way to interpret the mechanism behind their drive-noise tolerance is that by driving multiple modulators with the same signal, the same random error due to drive-noise is applied multiple times, such that it can be thought of as a coherent, systematic error.
The additional stages of the MZL can then be used to engineer some amount of cancellation of these systematic errors.
An issue with existing device designs \cite{Campenhout:09,VanCampenhout:11} is that they impart drive-noise dependent phase shifts (see appendix~\ref{IBM designs}).

In this work we extend the function of existing device designs to create devices which achieve drive-noise tolerance in both their on and off state.
Some of our proposed devices additionally have the property that they do not impart drive-noise dependent phase shifts on to the switched light modes.
Our designs are inspired by \textit{composite pulse sequences}, a technique originally created for correcting systematic errors which arise in Nuclear Magnetic Resonance.

\section{Composite pulses \label{composite pulses}}
Nuclear Magnetic Resonance (NMR), which is the study and control of nuclear spins within molecules using static and radio frequency magnetic fields~\cite{EBWbook}, has widespread applications in chemistry and biochemistry, and was also one of the first systems used for demonstrating quantum information processing (QIP)~\cite{Jones1998c,Chuang1998}.
A common problem arises from the radio frequency control fields, which are inhomogeneous over the ensemble of molecules in a macroscopic sample.
This creates a spatially varying error in the angle of rotation of the spins, leading to loss of information in the ensemble averaged signal.
A popular solution to this problem is to replace the single pulse used for a rotation with a composite pulse~\cite{LEVITT1979473}.
Appropriately engineering the angle and axis of rotation of each pulse can successfully reduce errors.
Since their first demonstration, a wide variety of composite pulse sequences have been developed, designed to correct for various types of errors~\cite{Levitt1986}, and have found application in NMR QIP~\cite{Cummins2003}. 

Composite pulse sequences perform unitary evolutions on a 2-dimensional complex vector space (e.g.\ a qubit), and can be applied to any other system described by such a vector space. 
Examples include superconducting circuits~\cite{Collin2004}, electron spin resonance~\cite{Morton2005a}, photon polarisation~\cite{Ardavan2007b}, and atom interferometers~\cite{Dunning2014}. 
In this work, we extend these applications to optical switching, which we describe by the 2-dimensional complex vector space of the 2 electromagnetic field spatial modes of our switch. 

In NMR, composite pulses are also frequently used to tackle off-resonance errors~\cite{Jones1998c}, which causes a tilt in the rotation axis, as well as rotation angle errors, but in optical switching the rotation induced by a modulator is confined to the \(z\)-axis and so tilt errors are not a significant problem. 

\section{Device model \label{model}}
For simplicity, and to stay platform agnostic, we use an idealised model. All devices discussed in this work are built from directional couplers  \(\bm{C}(\gamma)\), fixed phase shifts \(\bm{P}(\phi)\), and modulators \(\bm{M}(s, \theta)\), which we define as unitary matrices that act on the 2-dimensional complex space of the amplitudes of the electromagnetic field modes
\begin{equation}
\bm{C}(\gamma) = \begin{pmatrix}
\cos\gamma/2 & -{\mathrm i}\sin\gamma/2 \\
-{\mathrm i}\sin\gamma/2 & \cos\gamma/2 
\end{pmatrix},
\end{equation}
\begin{equation}
\bm{P}(\phi) = \begin{pmatrix}
\exp({-\mathrm i} \phi/2) & 0 \\
0 & \exp({\mathrm i} \phi/2)
\end{pmatrix},
\end{equation}
\begin{equation}
\bm{M}(s, \theta) = \begin{pmatrix}
1 & 0 \\
0 & \exp({\mathrm i} \theta s)
\end{pmatrix}.
\label{modulator}
\end{equation}
The coupler is described by an angle, \(\gamma\), which can be controlled through the physical parameters for each instance of the device. 
In integrated directional couplers, these parameters are typically the strength of the evanescent coupling between waveguides and the length of the coupling region. 

A fixed phase shift of angle \(\phi\), is implemented by creating a difference between the two optical path lengths of the two optical modes.
This could be implemented using a slow precise modulator, for example using the thermo-optic effect.

We assume that all the fast electro-optic modulators within an MZL switch are controlled by the same drive signal.
This allows us to model the driving signal using a single parameter, \(s\), which we assume takes the same value for all the modulators at any given point in time. 
We define our devices such that \(s = 0\) describes the ideal off state and \(s = 1\) describes the ideal on state, and will consider the effects of small deviations from these ideal values.
The on state phase shift angle of each modulator, \(\theta\), can be controlled by changing the length of the modulator. To ensure the validity of this model, it is important that any other noise sources have a much smaller effect than the noise from the drive signal. 

For later sections, we will also use the Hadamard transformation, which we can express, up to a global phase, as
\begin{equation}
\bm{H} = 
\bm{P}\left(\frac{\pi}{2}\right) \bm{C}\left(\frac{\pi}{2} \right) \bm{P}\left(\frac{\pi}{2}\right) 
= \frac{1}{\sqrt{2}} 
\begin{pmatrix}
1 & 1 \\
1 & -1
\end{pmatrix}.
\end{equation}
With these definitions, we can define an MZL switch by a sequence of these transformations. 
Our target transfer matrices which we will use as comparison for our fidelity measures are
\begin{equation}
V_{\textit{off}} = 
\begin{pmatrix}
1 & 0 \\
0 & 1
\end{pmatrix},\quad
V_{\textit{on}} = 
\begin{pmatrix}
0 & 1 \\
1 & 0
\end{pmatrix}.
\label{eq:on}
\end{equation}
For later comparisons, we define an MZI as
\begin{equation}
\label{MZI1}
\bm{H}   \bm{M}\left(s,\pi \right)\bm{H} .
\end{equation}

We only consider errors due to drive-noise, which we model as an imperfect value of \(s\), meaning that we assume a linear response of our modulators and no imperfections in the passive optical devices. 
Although fabrication errors will introduce errors in passive optical devices, near perfect passive optical devices can be achieved for example by replacing each coupler with a Mach--Zehnder interferometer~\cite{ntp-thesis:12, Miller:15}, then tuning it to implement the appropriate transformation \(\bm{C}(\gamma)\) using either a reprogrammable~\cite{Wilkes:16} or \textit{set-and-forget}~\cite{Chen:17,7537593} phase shift. 
Any imbalanced losses in the devices will affect fidelity, so losses at each stage should be engineered to be balanced for any realization of these devices.
Errors due to other effects such as dispersion or nonlinear effects are beyond the scope of this work.

To assess the performance of our devices, we use two different fidelity measures on the transfer matrices of our devices, \(U\), against a target transfer matrix, \(V\). 
One is a unitary fidelity measure \cite{PhysRevA.67.012317}
\begin{equation}
F = \frac{|\Tr (U V^\dagger)|^2}{4}
\end{equation}
which is sensitive to errors in both intensity and phase.  The other is a purely intensity dependent fidelity \cite{Carolan711}
\begin{equation}
F_I = \frac{\Tr\left(|U V^\dagger|^2\right)}{2}
\end{equation}
which is insensitive to errors in phase. Here \(|A|^2\) indicates the elementwise absolute value squared of a matrix \(A\).
Determining which fidelity measure should be considered will depend on the application of interest.

\section{Mapping composite pulses to optical switches \label{mapping}}Composite pulses are described by a list of pairs of angles describing the angle of rotation, \(\theta\), and the azimuthal angle of the axis of rotation in the \(xy\) plane, \(\phi\). Each pulse in the list is typically formatted with the angle of the axis as a subscript to the angle of rotation
\begin{equation}
    \theta_\phi.
\label{pulse}
\end{equation}

A subset of composite pulses known as \textit{inversion pulses} are interesting in the context of optical switching as their operation corresponds to a switch in the on state as defined in equation~\ref{eq:on}.
As an example, the Levitt composite inversion pulse~\cite{LEVITT1979473}, which we use in section~\ref{on} is written as
\begin{equation}
    \frac{\pi}{2}_{\frac{\pi}{2}}\,\pi_0\,\frac{\pi}{2}_{\frac{\pi}{2}},
\label{tycko}
\end{equation}
where the angles are expressed in radians. The tolerance of rotation angle errors seen for the Levitt pulse is analogous to the refocussing of inhomogeneous broadening by spin echoes \cite{Hahn1950} and photon echoes \cite{AKH1966}.

In the photonic devices described in section~\ref{model}, our modulators are only able to perform \(z\)-axis rotations in the Bloch sphere.
To translate this \(z\)-axis rotation to an \(x\)-axis rotation, we place Hadamard transformations either side of our modulators, 
and to control the axis of rotation in the \(xy\) plane, we tilt this axis with fixed phase shifts.
Finally we note that the modulator in effect performs a rotation through angle $s\theta$, and
thus equation~\ref{pulse} becomes

\begin{equation}
    \bm{P}\left(\phi \right) \bm{H} \bm{M}\left(s, \theta \right) \bm{H} \bm{P}\left(-\phi \right)
\label{mzl_stage}
\end{equation}

\begin{figure*}
    \centering
    \includegraphics[width=0.7\textwidth]{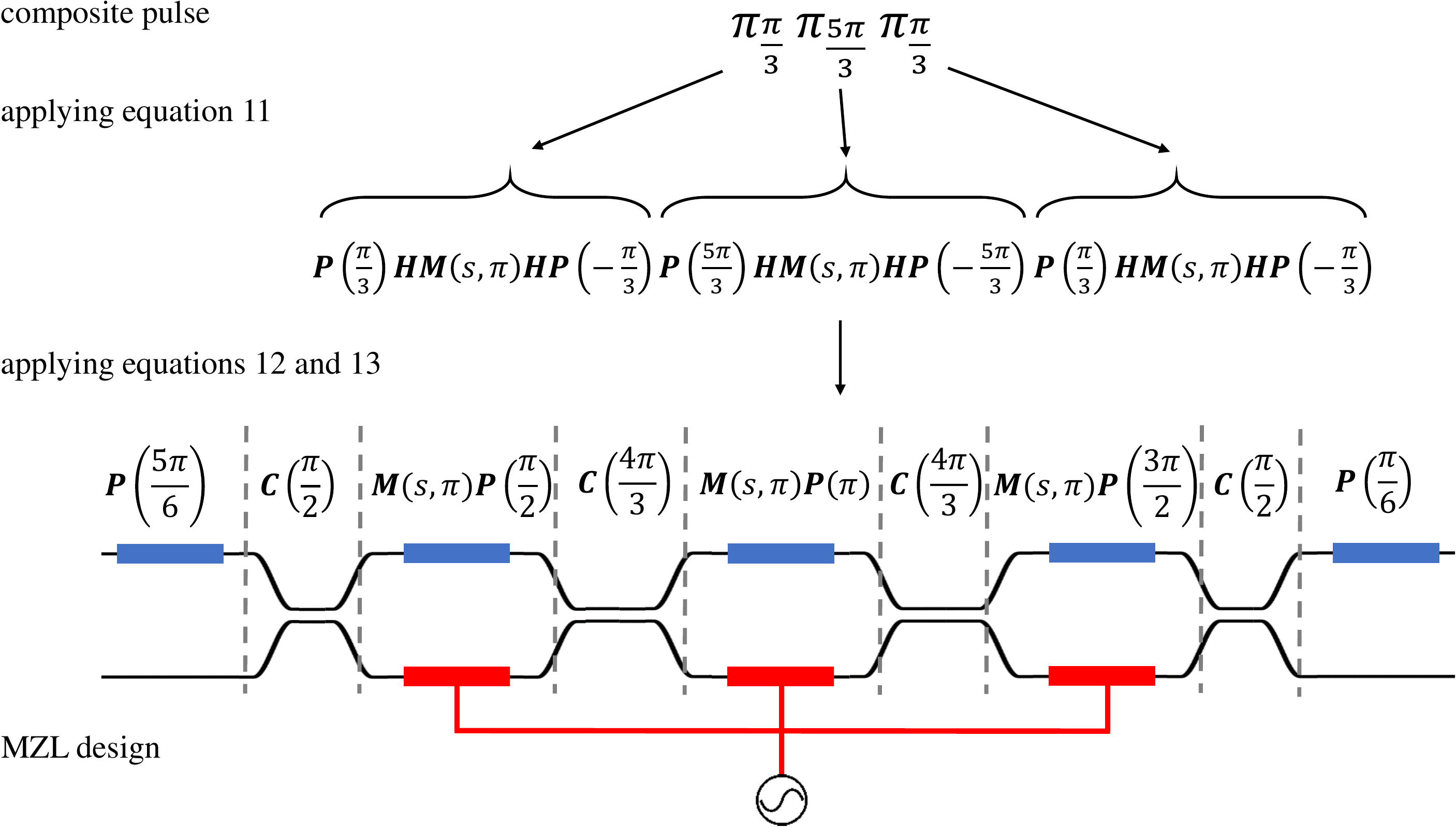}
    \caption{Outline of procedure for mapping a Tycko composite pulse~\cite{PhysRevLett.51.775} to an MZL switch design. Red rectangles represent the modulators, \(\bm{M\left(s, \theta \right)}\), and blue rectangles represent static phase shifts, \(\bm{P} \left( \phi \right) \). If we ignore global phases, the static phase shifts \(\bm{P} \left( \phi \right) \) can be implemented by acting a phase shift on only the upper mode.}
    \label{constructing_mzl}
\end{figure*}

By implementing this mapping from equation~\ref{pulse} to equation~\ref{mzl_stage} for each pulse in a sequence, we can find designs for composite pulse inspired MZL switches. These designs can then be simplified by applying the following identities:

\begin{equation}
    \bm{P}\left(a \right) \bm{P}\left(b \right) \equiv \bm{P}\left(a + b \right)
\label{mzl_identity1}
\end{equation}

\begin{equation}
    \bm{H} \bm{P} \left( \phi \right) \bm{H} \equiv \bm{C}\left(\phi \right) \equiv \bm{P} \left(\pi \right) \bm{C} \left(-\phi \right) \bm{P} \left(\pi \right)
\label{mzl_identity2}
\end{equation}
In figure~\ref{constructing_mzl}, we show an example of applying this mapping for constructing an MZL design using the Tycko composite pulse~\cite{PhysRevLett.51.775}. There is some flexibility when applying the identities of equations \ref{mzl_identity1} and \ref{mzl_identity2}, so the designer should consider what parameters they wish to optimise for. For robustness to fabrication error, each coupler, \(\bm{C}\), should be replaced by two couplers and a fixed phase shift as proposed in~\cite{ntp-thesis:12, Miller:15}.

In the next section, we look at the properties of MZL switches which have been designed using this mapping.

\section{MZL designs from known composite pulses \label{composite pulse mzls}}

In this section, we show how mapping different composite inversion pulses to MZL switches can achieve switching with a range of different regimes of drive-noise tolerance.
An important metric we will use to assess our devices is how successfully these composite pulse sequences can remove errors in the Taylor series expansion of the fidelity.
Composite pulses can remove some of the low order terms which take the fidelity away from unity.
If we first look at a conventional MZI, as defined by equation~\ref{MZI1}, we get the same expression for both our fidelity measures. The Taylor series  of fidelity against the ideal on state around \(s=1\) gives
\begin{equation}
    F = F_I = 1 - \varepsilon^2 - O\left( \varepsilon^4 \right),
\end{equation}
where we have defined \(\varepsilon = \pi (s - 1) / 2\). When we look at off state fidelity we instead use \(\varepsilon = \pi s / 2\).
If we compare this to the Levitt composite NOT pulse sequence~\cite{LEVITT1979473}, the first known composite pulse, we find that \(F\) is the same as for the MZI, but the intensity only fidelity measure for the on state gives
\begin{equation}
    F_I = 1 - \varepsilon^4 - O\left( \varepsilon^6 \right).
\end{equation}
So for the Levitt pulse, we can say that the leading term in the expansion in the on state of \(F\) is \(\varepsilon^2\) and for \(F_I\) is \(\varepsilon^4\).

Another important parameter for us to keep track of is the \textit{modulation depth}. This quantity is proportional to the total length of the modulators that the light will propagate through. We define modulation depth as
\begin{equation}
    D = \frac{1}{\pi}\sum_i \theta_i
\end{equation}
where \(\theta_i\) gives the length of the \(i\)th modulator as defined in equation~\ref{modulator}, which means that \(D = 1\) for an MZI.

A description of the Levitt pulse and all other pulses in this text as both composite pulse sequences and MZL switches can be found in appendix~\ref{sequences and mzls}.

\subsection{On state drive-noise tolerance \label{on}}
\begin{figure}
    \centering
    \includegraphics[width=0.47\textwidth]{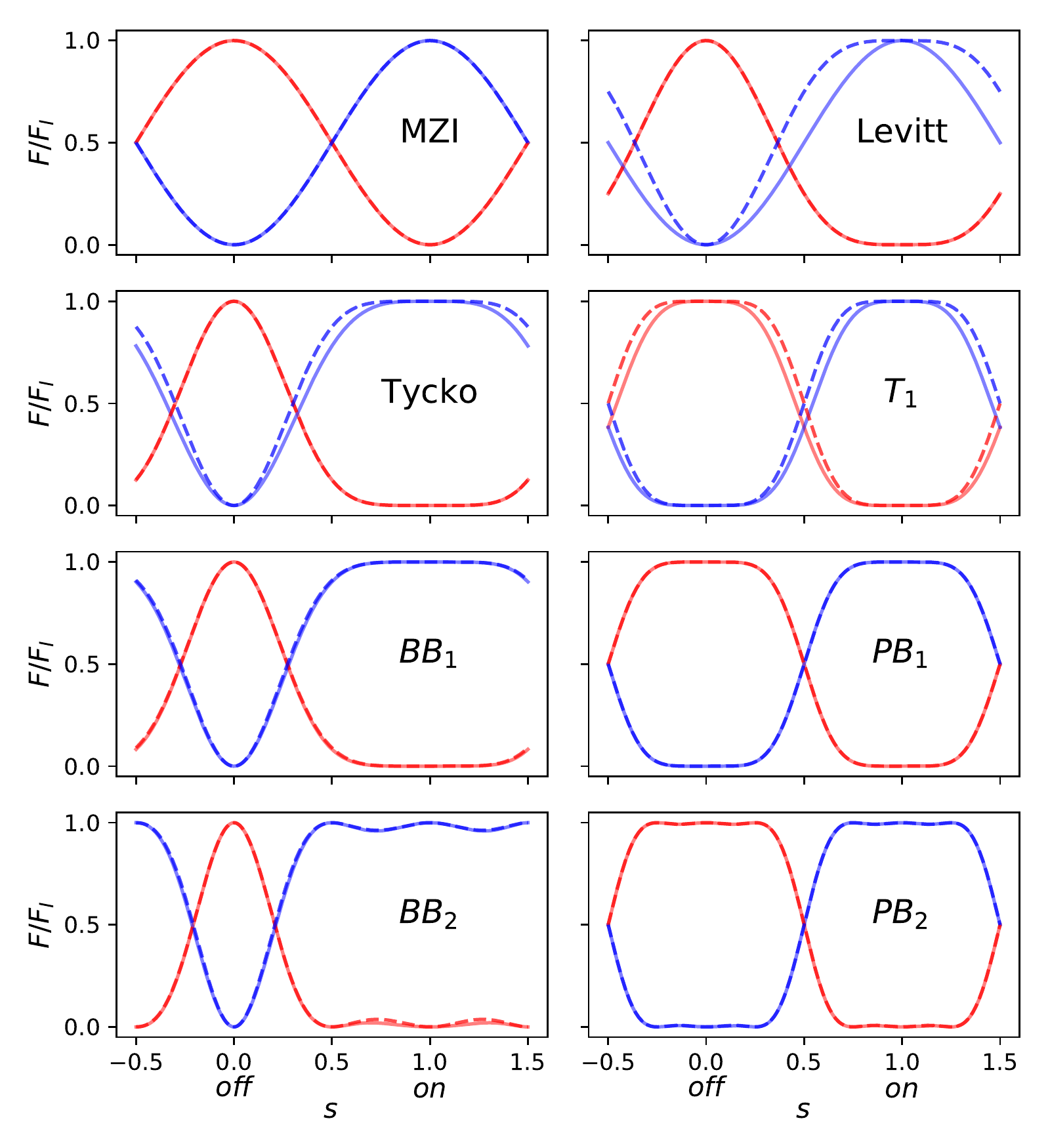}
    \caption{\(F\) (solid) and \(F_I\) (dashed) plotted for all the optical switch designs discussed in section~\ref{composite pulse mzls} when compared to the ideal off state (red) and on state (blue).}
    \label{fidelities}
\end{figure}

\begin{table}
\centering
\begin{tabular}{l|l|ll|ll}
         &   &\multicolumn{4}{c}{leading infidelity term} \\ \cline{3-6}
Device  & D &\multicolumn{2}{c|}{on state}&\multicolumn{2}{c}{off state}\\ 
  &  & \(F\) & \(F_I\) & \(F\) & \(F_I\)\\ \hline
MZI      & 1 & $\varepsilon^2$ & $\varepsilon^2$ & $\varepsilon^2$ & $\varepsilon^2$         \\
Levitt   & 2 & $\varepsilon^2$ & $\varepsilon^4*$ & $2\varepsilon^2$ & $2\varepsilon^2$          \\
Tycko    & 3 & $0.75\varepsilon^4*$ & $\varepsilon^6*$ & $3\varepsilon^2$ & $3\varepsilon^2$          \\
T\(_1\)  & 5 & $3.75\varepsilon^4$ & $10\varepsilon^6$ & $3.75\varepsilon^4*$ & $10\varepsilon^6*$     \\
BB\(_1\) & 5 & $0.625\varepsilon^6*$ & $0.625\varepsilon^6$ & $3.75\varepsilon^2$ & $3.75\varepsilon^2$ \\
PB\(_1\) & 9 & $7.875\varepsilon^6$ & $7.875\varepsilon^6$ & $7.875\varepsilon^6*$ & $7.875\varepsilon^6$ 
\end{tabular}
\caption{Performance summary for devices presented in section~\ref{composite pulse mzls}. An asterisk (*) is used to indicate when a pulse out-performs any pulse with smaller depth D on the measure indicated.} \label{device performance table}
\end{table}

The Levitt pulse, discussed above, achieves improved drive-noise tolerance when measured by \(F_I\) and has an optical depth, \(D=2\). By going to larger modulation depths, we can further improve our drive-noise tolerance. For \(D=3\), the Tycko pulse~\cite{PhysRevLett.51.775} offers on state drive-noise tolerance in both \(F\) and \(F_I\) fidelity measures, with a leading infidelity term of \(0.75\varepsilon^4\) for \(F\) and \(\varepsilon^6\) for \(F_I\).

\subsection{On and off state drive-noise tolerance \label{on off}}

At a modulation depth of \(D = 5\), there exists an interesting composite pulse sequence, designed by Wimperis, which we refer to as T\(_1\)~\cite{WIMPERIS1989509}.
This pulse sequence is the shortest known example which is designed to have a rectangular profile, meaning that when we use it in an MZL switch, we have drive-noise tolerance in both the on and off states.
The T\(_1\) pulse has a leading infidelity term of \(3.75\varepsilon^4\) for \(F\) and \(10\varepsilon^6\) for \(F_I\), but unlike before, this is true in both the on and off state, as shown in Figure~\ref{fidelities}.
Achieving both on and off state drive-noise tolerance is particularly important for \textit{push-pull} modulation schemes. 
Push-pull refers to when there is a modulator on both of the paths through the interferometer and the phase shift is applied by alternating which modulator is switched on, meaning that voltage needs to be applied to the device in both the on and off states.
\subsection{Increasing fidelity}
To find an MZL design which can suppress the 4th order unitary infidelity terms for the on state, we again need to go up to a modulation depth, \(D = 5\). The \textit{broadband} pulse BB\(_1\)~\cite{WIMPERIS1994221}, designed by Wimperis, meets this criterion.

To find a pulse sequence which can suppress 4th order infidelity terms of \(F\) in both the on and off state, we look to the \textit{passband} pulses by Wimperis~\cite{WIMPERIS1994221}. The PB\(_1\) pulse requires a modulation depth of \(D=9\) and suppresses 2nd and 4th order terms of infidelity for \(F\) and \(F_I\) in both on and off states. The devices presented so far are summarised in Table~\ref{device performance table}.

\subsection{Trading fidelity for increased tolerance}
The pulses presented in sections~\ref{on} and~\ref{on off} have been optimised for maximum fidelity in the presence of small errors due to drive-noise. However, other pulses exist which offer protection against still larger errors, but at a slightly lower fidelity for small errors. Wimperis explored this trade off and created the BB\(_2\) and PB\(_2\) pulse~\cite{WIMPERIS1994221}, which are related to the BB\(_1\) and PB\(_1\) pulses. These pulses are shown in the bottom row of Figure~\ref{fidelities}.

\section{Discussion}
By leveraging the mature field of composite pulses, we have shown that new regimes of drive-noise tolerance in optical switches should be possible. We present designs which achieve drive-noise tolerance without decreasing unitary fidelity and designs which show on and off state drive-noise tolerance, a goal set out in~\cite{Campenhout:09}. This enables high fidelity push-pull based optical switches even with a noisy drive signal. Whilst for some applications, the cost of increased complexity of using MZL switches will not justify the improvements to drive noise tolerance, we have identified several areas where drive noise tolerance is critically important and so this trade off will be worthwhile. We also believe that there may be other ways to leverage composite pulses for designing other integrated photonics devices, such as MZL broadband couplers~\cite{Tormen:05, 82964,202833, Lu:15} or filters~\cite{350600,Horst:13, doi:10.1002/lpor.201200032}.

All of the devices discussed here have been designed based on inversion pulses, but it may be interesting to look at composite pulses for other rotation angles.
These might find application, for example, in drive-noise tolerant adaptive measurements of dual rail photonic qubits. The BB\(_1\) and PB\(_1\) pulses are easily adapted for other rotation angles~\cite{WIMPERIS1994221}, and so provide a completely general solution to this problem. 

It is possible to continue suppressing further terms in the infidelity in the on state by designing MZL switches with higher modulation depth~\cite{PhysRevA.70.052318,JONES20132860}. However, we think that this trade off will not be practical due to the increased losses, footprint and power consumption associated with higher modulation depth. 
Also, composite pulses are designed to tackle systematic errors where the general form of the errors, but not their precise magnitude, is known, and so only work well when the dominant errors take the expected form.
Simulations and experimental studies in the context of NMR~\cite{PhysRevA.73.032334, PhysRevA.90.012316, zhen2016experimental, PhysRevA.100.023410} suggest that composite pulses are an effective means of error suppression in many cases, but that short and simple sequences may in practice work better than more complex sequences with theoretically better performance.
Although fabrication errors can be suppressed near perfectly using the techniques of~\cite{Miller:15, ntp-thesis:12}, remaining effects due to fabrication error and error from other effects will limit the benefit of using the more complex MZL switches.

Challenges remain to realise devices which can fully capitalise on these advances. 
Using more realistic and platform specific models to inform fine tuning of the device designs will help to offset issues such as fabrication imperfection and loss.
The designs presented require precise control over many internal phase shifts within the device.
If this is to be achieved using tuning or trimming, it presents a calibration challenge.
It may be required to monitor transmission of light at each stage of the MZL, through a process similar to proposals for calibrating passive MZL filters~\cite{Miller:17}.
Options to achieve this include weakly coupled integrated detectors, ``taps'' to route some of the light to external detectors or frequency selective Bragg gratings~\cite{Calkins:13}.
To avoid increased losses, these taps could be made using a process which allows for them to be removed~\cite{trimming}.
Alternatively, a modular construction~\cite{wc-thesis:18, Mennea:18} would allow for each stage to be calibrated individually.

\acknowledgements{
Engineering and Physical Sciences Research Council (EPSRC) (EP/N509711/1)
}


\begin{appendices}

\section{Existing drive-noise tolerant MZL designs \label{IBM designs}}
\begin{figure}
    \centering
    \includegraphics[width=0.4\textwidth]{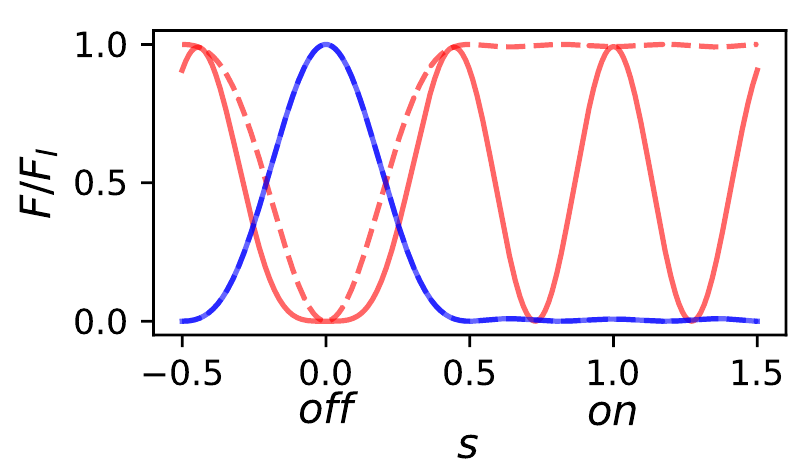}
    \caption{\(F\) (solid) and \(F_I\) (dashed) for MZL used in~\cite{VanCampenhout:11}, and designed according to the methods in~\cite{Campenhout:09}, when compared to the ideal off state (red) and on state (blue). Ideal on and off states for this MZL are the reverse of the convention outlined in section~\ref{model}.}
    \label{ibm fig}
\end{figure}

The designs in~\cite{Campenhout:09} were created by imagining taking an MZI and breaking up the modulation into $N$ smaller pieces, surrounded by $N+1$ couplers.
To ensure extinction in the off state, their couplers are constrained such that
\begin{equation}
\sum_{i=1}^{N + 1} \gamma_i = \pi
\end{equation}
and they chose to make their devices symmetric i.e. \(\gamma_i = \gamma_{N+1-i}\).
They break up the modulator evenly, i.e. \(\theta_i = \pi/N \).
With these shorter modulators, they found that they needed to apply larger voltages to reach the on state than for a MZI. If we rescale our \(s\) parameter and \(\theta_i\) parameters such that the on state occurs at \(s=1\), we find that this gives \(\theta_i = \pi\) for all devices presented in~\cite{Campenhout:09}.
Figure~\ref{ibm fig} shows the performance of the design from~\cite{VanCampenhout:11}, which was made using the methods in~\cite{Campenhout:09}, when modelled in the framework used in this work.
As seen in the figure, the fidelity of these devices depends strongly on whether a phase sensitive or phase insensitive fidelity measure is used, and therefore renders these devices unsuitable for applications where phase information is important.

It is worth commenting that our simplistic device model succeeds in describing the drive-noise tolerance behaviour of devices which have been experimentally realised~\cite{VanCampenhout:11}. We expect that the reverse will be true, and that the devices we present in this work will inspire real devices with enhanced drive-noise tolerance, despite our model excluding many forms of imperfections.

\section{Composite pulses and Mach--Zehnder Lattice designs \label{sequences and mzls}}

Table~\ref{sequences} shows the result of using the mapping and simplifications described in section~\ref{mapping}, applied to the sequences discussed in this paper. Simplifications were applied to reduce the terms in the sequence, rather than to suggest the most practically appropriate design. We let \(\alpha = \arccos(-1/4)\) and \(\beta = \arccos(-1/8)\).

\begin{table*}
\caption{Composite pulse sequences and MZL switch designs for all devices discussed \label{sequences}}
\begin{tabular}{l|l|l}
Design   & Pulse sequence                                                               & MZL switch design                                                                                                                                                                                                                                                                                       \\ \hline
MZI      & \(\pi_0\)                                                                    & \(\bm{H} \bm{M} \left(s, \pi \right) \bm{H}\)                                                                                                                                                                                                                                             \\
Levitt   & \(\frac{\pi}{2}_\frac{\pi}{2} \pi_0 \frac{\pi}{2}_\frac{\pi}{2}\)            & \(\bm{P} \left( \frac{\pi}{2} \right) \bm{H} \bm{M} \left(s, \frac{\pi}{2} \right) \bm{C} \left( -\frac{\pi}{2} \right) \bm{M} \left(s, \pi \right) \bm{C} \left( \frac{\pi}{2} \right) \bm{M} \left(s, \frac{\pi}{2} \right) \bm{H} \bm{P} \left(-\frac{\pi}{2} \right) \)               \\
Tycko    & \( \pi_{\frac{\pi}{3}} \pi_{\frac{5\pi}{3}} \pi_{\frac{\pi}{3}} \)           & \( \bm{P} \left( \frac{\pi}{3} \right) \bm{H} \bm{M} \left(s, \pi \right) \bm{C} \left( \frac{4\pi}{3} \right) \bm{M} \left(s, \pi \right) \bm{C} \left( -\frac{4\pi}{3} \right) \bm{M} \left(s, \pi \right) \bm{H} \bm{P} \left( -\frac{\pi}{3} \right)\)                                \\
T\(_1\)  & \(2\pi_{-\alpha} 2\pi_{\alpha} \pi_0 \)                                      & \(\bm{P} \left(\alpha \right) \bm{H} \bm{M} \left( s, 2\pi \right)  \bm{C} \left( 2\alpha \right) \bm{M} \left(s, 2\pi \right) \bm{C} \left(-\alpha \right) \bm{M} \left( s, \pi \right) \bm{H}\)                                                                                         \\
BB\(_1\) & \( \pi_{\alpha} 2\pi_{3\alpha} \pi_{\alpha} \pi_0 \)                        & \(\bm{P} \left(\alpha \right) \bm{H} \bm{M} \left(s, \pi \right) \bm{C} \left( 2\alpha \right) \bm{M} \left( s, 2\pi \right) \bm{C} \left(-2\alpha \right) \bm{M} \left(s, \pi \right) \bm{C} \left(-\alpha \right) \bm{M} \left(s, \pi \right) \bm{H} \) \\
PB\(_1\) & \(2\pi_{\beta} 4\pi_{-\beta} 2\pi_{\beta} \pi_0\)                            & \( \bm{P} \left( \beta \right) \bm{H} \bm{M} \left(s, 2\pi \right) \bm{C} \left( -2\beta \right) \bm{M} \left(s, 4\pi \right) \bm{C} \left( 2\beta \right) \bm{M} \left(s, 2\pi \right) \bm{C} \left( -\beta \right) \bm{M} \left(s, \pi \right) \bm{H} \)\\
BB\(_2\) & \( \pi_{\frac{\pi}{2}} 2\pi_{-\frac{\pi}{4}} \pi_{\frac{\pi}{2}} \pi_0 \)    & \( \bm{P} \left( \frac{\pi}{2} \right) \bm{H} \bm{M} \left(s, \pi \right) \bm{C} \left( -\frac{3\pi}{4} \right) \bm{M} \left(s, 2\pi \right) \bm{C} \left( \frac{3\pi}{4} \right) \bm{M} \left(s, \pi \right) \bm{C} \left( -\frac{\pi}{2} \right) \bm{M} \left(s, \pi \right) \bm{H}\)  \\
PB\(_2\) & \( 2\pi_{\frac{\pi}{2}} 4\pi_{-\frac{5\pi}{8}} 2\pi_{\frac{\pi}{2}} \pi_0 \) & \( \bm{P} \left( \frac{\pi}{2} \right) \bm{H} \bm{M} \left(s, 2\pi \right) \bm{C} \left( \frac{7\pi}{8} \right) \bm{M} \left(s, 4\pi \right) \bm{C} \left( -\frac{7\pi}{8} \right) \bm{M} \left(s, 2\pi \right)  \bm{C} \left( -\frac{\pi}{2} \right) \bm{M} \left(s, \pi \right) \bm{H}\) 

\end{tabular}
\end{table*}

\end{appendices}

\newpage
\bibliographystyle{apsrev4-1}
\bibliography{refs.bib}

\end{document}